\documentclass[aps,prd,showpacs,preprintnumbers]{revtex4}
\usepackage{amsmath,amssymb}
\usepackage{graphicx}

\newcommand\bef{\begin{figure}}
\newcommand\eef[1]{\label{fg:#1}\end{figure}}
\newcommand\beq{\begin{equation}}
\newcommand\eeq[1]{\label{#1}\end{equation}}
\newcommand\beqa{\begin{eqnarray}}
\newcommand\eeqa[1]{\label{#1}\end{eqnarray}}
\newcommand\bet{\begin{table}}
\newcommand\eet[1]{\label{tb:#1}\end{table}}

\newcommand\fgn[1]{Figure \ref{fg:#1}}
\newcommand\eqn[1]{eq.\ (\ref{#1})}
\newcommand\scn[1]{Section \ref{sec:#1}}

\newcommand\tbn[1]{Table \ref{tb:#1}}

\newcommand\etal{{\sl et al.\/}}

\newcommand{\MSbar}{{\overline{\scriptscriptstyle MS}}}

\newcommand\ket[1]{\left|{#1}\right\rangle}
\newcommand\bra[1]{\left\langle{#1}\right|}

\begin{document}
\title{Statistical tweaks and flow scale from masses}
\author{Saumen\ \surname{Datta}}
\email{saumen@theory.tifr.res.in}
\affiliation{Department of Theoretical Physics, Tata Institute of Fundamental
         Research,\\ Homi Bhabha Road, Mumbai 400005, India.}
\author{Sourendu\ \surname{Gupta}}
\email{sgupta@theory.tifr.res.in}
\affiliation{Department of Theoretical Physics, Tata Institute of Fundamental
         Research,\\ Homi Bhabha Road, Mumbai 400005, India.}
\author{Anirban\ \surname{Lahiri}}
\email{anirban@theory.tifr.res.in}
\affiliation{Department of Theoretical Physics, Tata Institute of Fundamental
         Research,\\ Homi Bhabha Road, Mumbai 400005, India.}
\author{Pushan\ \surname{Majumdar}}
\email{tppm@iacs.res.in}
\affiliation{Department of Theoretical Physics, Indian Association for the
         Cultivation of Science,\\ Raja Subodh Chandra Mallick Road,
         Jadavpur, Kolkata 700032, India.}
\begin{abstract}
We compare lattice scales determined from the vector meson mass and
the Wilson flow scale $w_0$ in QCD with two-flavours of rooted naive
staggered fermions over a wide range of lattice spacing and quark
mass. We find that the distributions of meson correlation functions
are non-Gaussian. We modify the statistical analysis to take care
of the non-Gaussianity. Current day improvements in the statistical
quality of data on hadron correlations further allow us to simplify
certain aspects of the analysis of masses. We examine these changes
through the analysis of pions and apply them to the vector meson. We
compare the flow scale determined using the rho mass with that using
the $\Lambda_{\overline{\scriptscriptstyle MS}}$.
\end{abstract}
\preprint{TIFR/TH/16-20}

\maketitle

\section{Introduction}

In any lattice computation, one needs to specify the lattice scale. The
earliest approach to this was to determine a hadron mass on the lattice
and use this to set the scale. The difficulty of determining hadron
masses with small systematic errors has gradually led to development of
other techniques.  Currently the simplest seems to be the flow scale
$w_0$ \cite{sommer}.  Since this is a theory scale, not measurable in
experiments, it is useful to compare it with other scales, and with the
same scale determined using different lattice actions. Such comparisons
quantify the approach to the continuum limit.

In a previous work we examined the flow scale with two flavours of
rooted naive staggered quarks over a large range of lattice spacing
and quark masses \cite{prev}. By comparing the flow scale with
$\Lambda_{\overline{\scriptscriptstyle MS}}$, determined using the
Lepage-McKenzie scheme, we found that $w_0=0.13^{+0.01}_{-0.02}$ fm.
This flow scale is smaller than those obtained using other discretizations
of the Dirac operator. However, there may be UV corrections in comparing
the flow scale with the QCD scale so determined, which have not been
examined yet. So we examine here the ensembles generated in the earlier
study to determine the flow scale from the vector meson mass.

This leads us to re-examine the extraction of meson masses from naive
staggered quarks. The last such measurements were performed 25 years ago,
when the current technology of using covariance matrices in fits was
just a few years old \cite{covmat}. However, computational hardware has
scaled from a few hundred megaflops to a few hundred teraflops in the
intervening years. As a result, one can beat down auto-correlations
between configurations tremendously even with the old algorithms.
With a set of almost uncorrelated configurations one may use simpler
statistical tools.

Consider one of the changes possible if the sampling of lattice gauge
configurations became cheap. Measurements of correlation functions
could be done using completely different sets of a very large number
of configurations at each distance, each drawn from a thermalized
configuration. Since the measurements of correlators at each separation
would then be statistically independent, the covariance between them
would vanish, and it would be easier to fit masses to them. While this
ideal is still out of reach, it is worth considering analyses which make
it simple to transit from statistics-limited to large-statistics studies.

Our basic tool is the widely used non-parametric method called
bootstrap or resampling \cite{efron}. For any statistic obtained from
a configuration, we estimate its distribution non-parametrically by
bootstrap.  This versatile construction replaces the assumption that the
statistic is Gaussian distributed.  The median and the 34\% limits above
and below it are taken as the non-parametric estimate of the average
and error. The sample median has the nice property that its distribution
tends to a Gaussian. This is an old theorem, whose proof we present in
the appendix, since it seems to be relatively poorly known in lattice
gauge theory \cite{borici}.  Some limits on the use of bootstraps were
explored in \cite{ourboot}.  Using this for multi-dimensional resampling,
or by nesting bootstraps, one can generate a variety of analysis tools.

We report hadron masses determined in simulations with two flavours of
staggered quarks using the configurations described in \cite{prev}.
Measurements of the mass of the pseudo-Goldstone pion were reported
earlier using techniques similar to those described in \cite{covmat}.
We revisit that measurement, and also report our estimates of the masses
of vector mesons.

In the next section we report on an exploration of various statistical
methods using staggered pseudo-Goldstone pion correlators. In the section
after that we report measurements of vector meson masses. We compare the
setting of scale using $m_\rho$ and $w_0$, and give an estimate of $w_0$
from $m_\rho$. Our main conclusions are collected in the fourth section.
In an appendix we describe a theorem on the distribution of sample
medians which is useful for bootstrap estimates of random variates which
are not Gaussian distributed.

\section{Pions}\label{sec:pion}

\bef
\begin{center}
\includegraphics[scale=0.7]{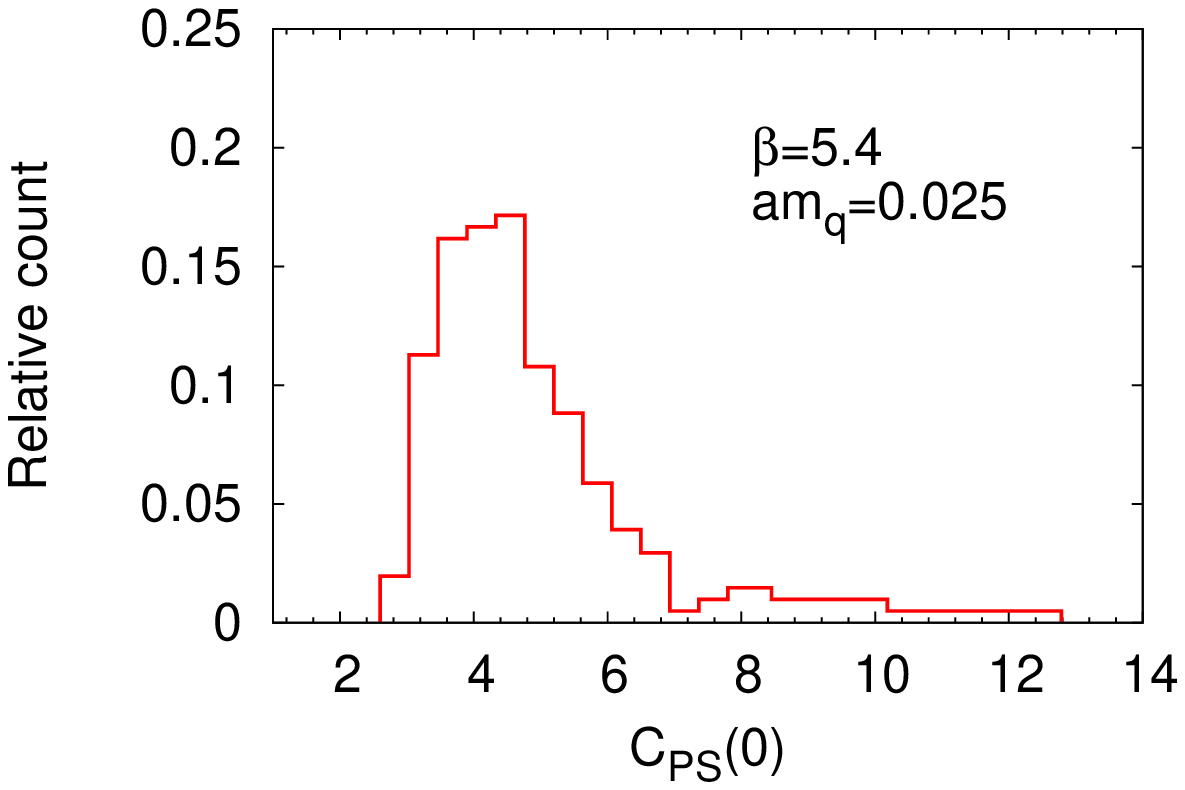}
\includegraphics[scale=0.7]{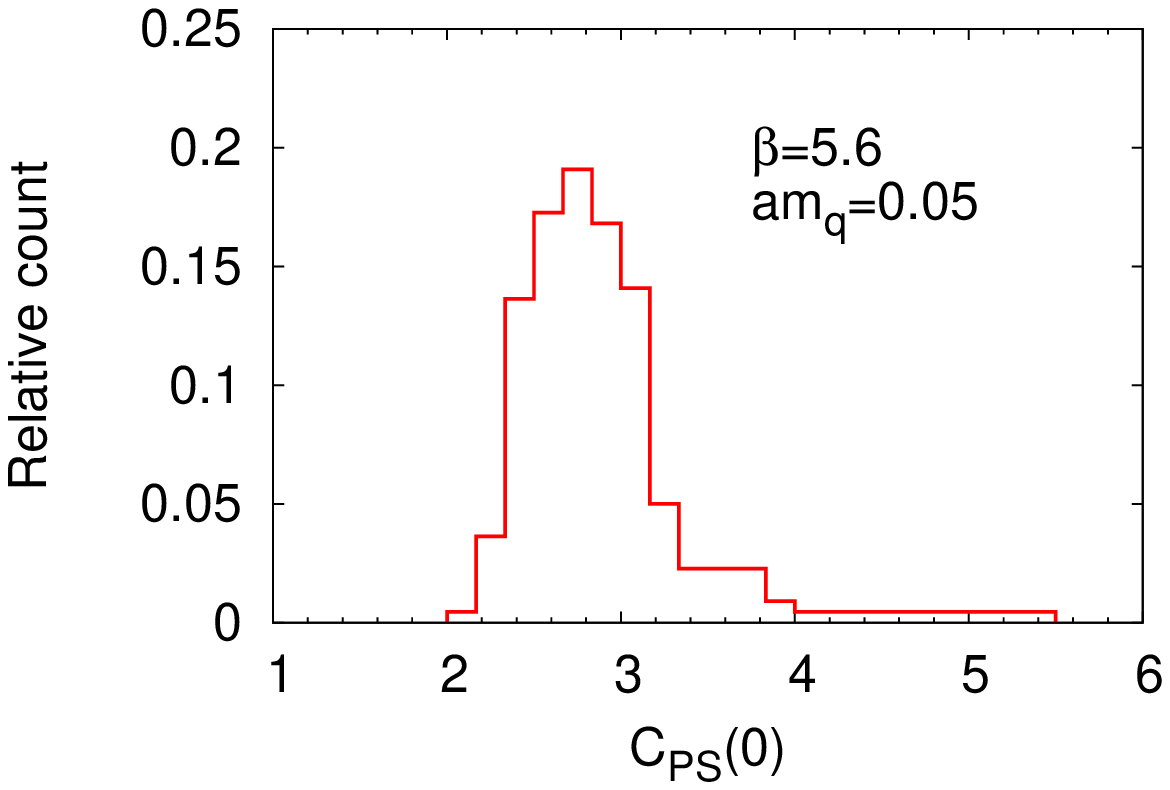}
\includegraphics[scale=0.7]{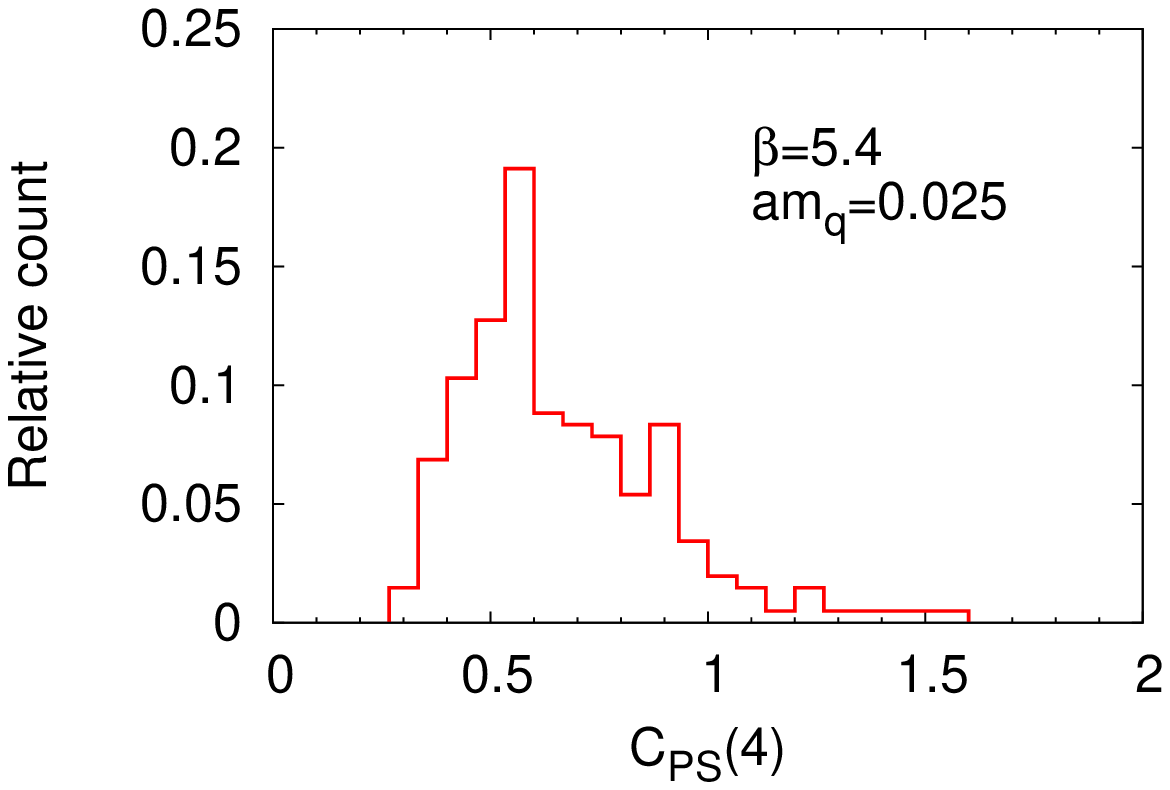}
\includegraphics[scale=0.7]{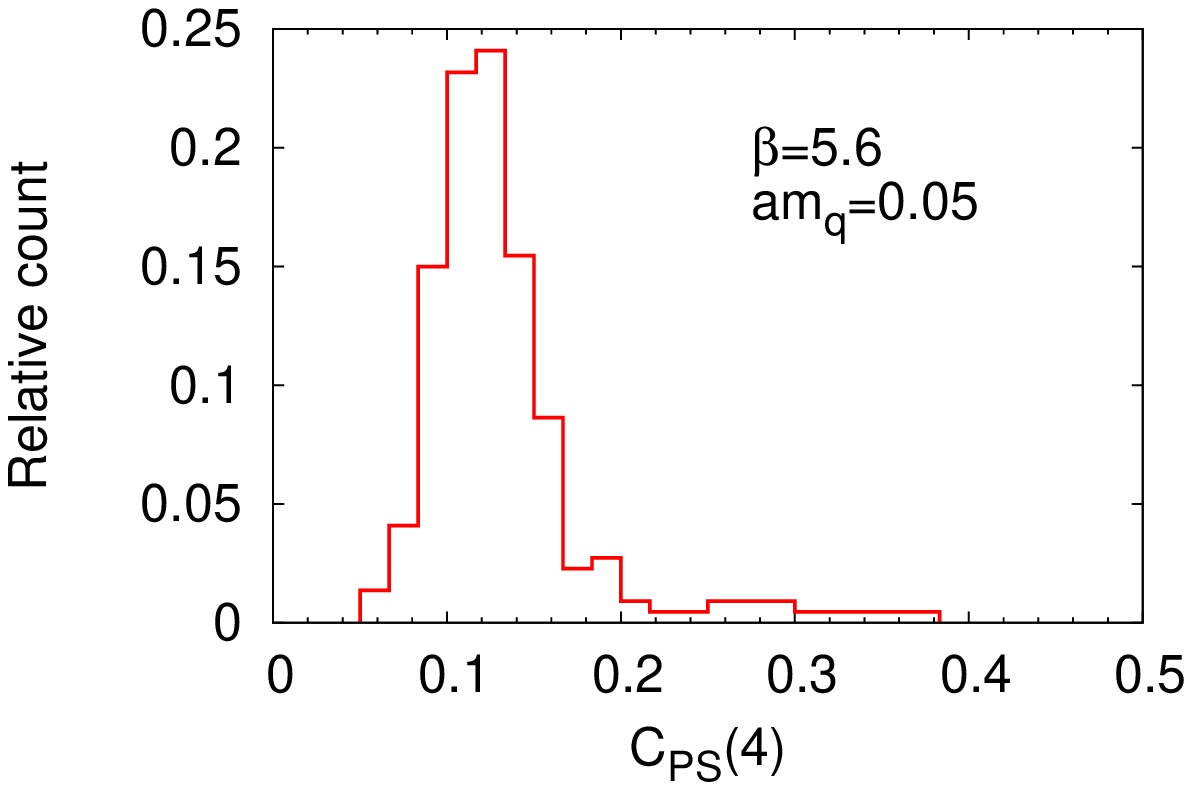}
\end{center}
\caption{The distribution of measurements of the PS correlator at distance
 zero for two representative run IDs 6 (left) and 14 (right). The ID numbers
 associated with runs are given in \tbn{psmass}. Such highly skewed
 distributions are seen to be generic.}
\eef{pscorrdist}

\bef
\begin{center}
\includegraphics[scale=1.0]{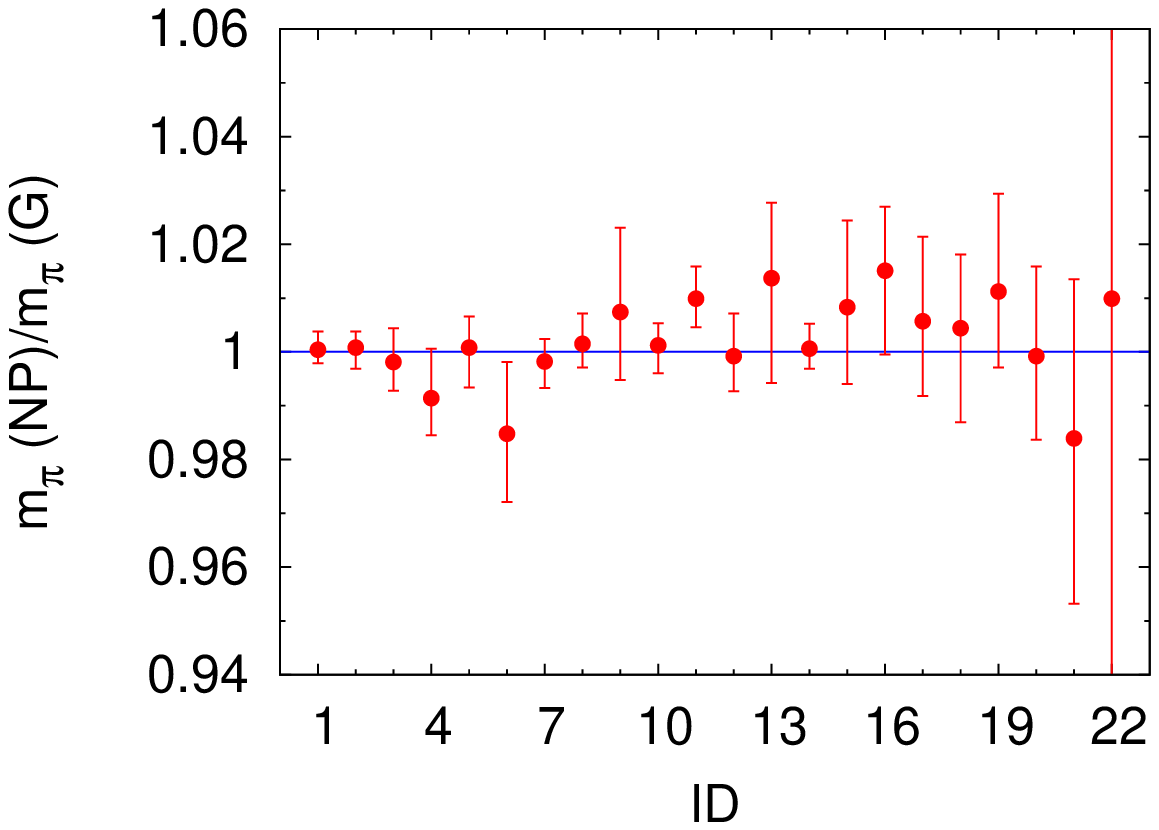}
\end{center}
\caption{The ratio of pion masses extracted with non-parametric bootstrap
 estimators of the correlator (labelled NP) and with Gaussian estimators
 (labelled G). The uncertainty in the ratio are obtained by repeating the
 bootstrap resampling and generating the distribution of the ratio. The ID
 numbers on the ordinate are associated with run parameters in \tbn{psmass}.}
\eef{psnpg}

The staggered pseudo-Goldstone pion is a good test-bed for exploring
statistical techniques for two reasons. First, because masses are typically
small, the relative uncertainty in the correlation function is small. Second,
unlike other staggered hadrons, there is no opposite parity channel to
complicate the analysis, and it is often sufficient to use a fitting form
\beq
   C_{PS}(t)=A\cosh\left[m\left(\frac{aN_t}2-t\right)\right],
\eeq{pionc}
where $a$ is the lattice spacing, and $A$ and $ma$ are fit parameters.

We begin by examining the distribution of the measurements of the
PS correlation function at fixed separation $t$.  Two representative
histograms of the measurements of the PS correlator are shown for each of
$t=0$ and $t=4$ in \fgn{pscorrdist}. They are highly skewed.  We found
that the set of configurations which give rise to measurements in the
tail of the distribution of the correlator at $t=0$ are generically
those which populate the tail at other $t$.  Such long-tailed and skewed
ensembles of the correlators are generic, in the sense that we saw such
distributions for all 22 cutoffs and quark masses which we examined.

This is not an artifact of a lack of thermalization. The sequence of
configurations which we used were thermalized according to global
measurements like average plaquette or quark condensate. Moreover,
the measurements which lie at the tails of the distribution are
distributed throughout the runs, and not clustered together. 

Non-Gaussian distributions of correlation functions have been sporadically
reported in the literature \cite{nong}. Unlike those, the distributions
which we see are not log-normal. In fact, in this case even the logarithms
of correlation functions have long-tailed distributions. An extreme
example is for ID number 14 (see \tbn{psmass} for the association of
ID numbers with run parameters). For this set the distribution of
$\log C_{PS}(4)$, shown in the last panel of \fgn{pscorrdist}, has
skewness $-2.5$ and kurtosis 25.  Systematic reports of distributions
of correlators is not part of the standard analysis suite of lattice
gauge theory.  As a result, we do not know whether our observation is
of greater generality.

We investigated the configurations which give measurements in the long
tail of the distribution. Pion correlation functions measured on these
configurations with different source locations do not have large values
of the correlator for all locations of the source. In order to understand
this, it is useful to think in terms of the eigenvalue decomposition of
the correlators.

Suppose that $\lambda_i$ is an eigenvalue of the staggered Dirac operator,
and the component of the eigenvector at site $x$ is $\ket{\lambda_i(x)}$,
then
\beq
   D_{xy}\ket{\lambda_i(y)}=\lambda_i\ket{\lambda_i(x)}, \qquad{\rm and}\qquad
   D^{-1}_{xy}=\sum_i\frac1{\lambda_i}\;\ket{\lambda_i(x)}\bra{\lambda_i(y)}.
\eeq{eigs}
As a result, any local mesonic correlation function is given by
\beq
   C(x,y)=\sum_{ij}\frac1{\lambda_i\lambda_j}
     \bra{\lambda_i(x)}\gamma\ket{\lambda_j(x)}
     \bra{\lambda_j(y)}\gamma\ket{\lambda_i(y)},
\eeq{corrfn}
where $\gamma$ is the spin-flavour matrix which enters the source for the
meson under consideration. If a configuration has one or more eigenvalues
$\lambda_i$ which are very small compared to the minimum eigenvalue in a
generic configuration, then the correlation function becomes large. If,
at the same time, the corresponding eigenvectors are localized, then
$C(x,y)$ may not be equally large for all $x$ and $y$. This would imply
that on these configurations some sources give much larger values of
the correlator than others. Previous investigations have shown that
the eigenvalues and eigenvectors have exactly this kind of behaviour
in the presence of topological structures \cite{topology}. This leads
us to believe that topology is generally the reason for the skewed
distributions which we see.\footnote{One other interesting conclusion
from \eqn{corrfn} is worth mentioning.  Interchanging $x$ and $y$ in the
meson correlator while simultaneously interchanging the dummy indices $i$
and $j$ shows that in each configuration $C(x,y)=C(y,x)$. Truncation of
the summation due to incomplete convergence of the Dirac operator does
not spoil this property,
nor does loss of arithmetic precision in one or more eigenvectors.}

Since the tail of the distribution of measurements of the correlators is
so long, it is not clear whether the central limit theorem applies.  As a
result, justification for the use of Gaussian statistics (labelled G),
through the use of means, variances, and covariance matrices, is lacking.
On the other hand, it is safe to use a non-parametric bootstrap analysis
(labelled NP).  This does have an effect on the determination of the
mass, as we show in \fgn{psnpg}. There is a tendency for NP to lead to
a higher mass, although, as the figure shows, the discrepancy between
the two methods is generally less than a 1-$\sigma$ effect.

\bef
\begin{center}
\includegraphics[scale=0.7]{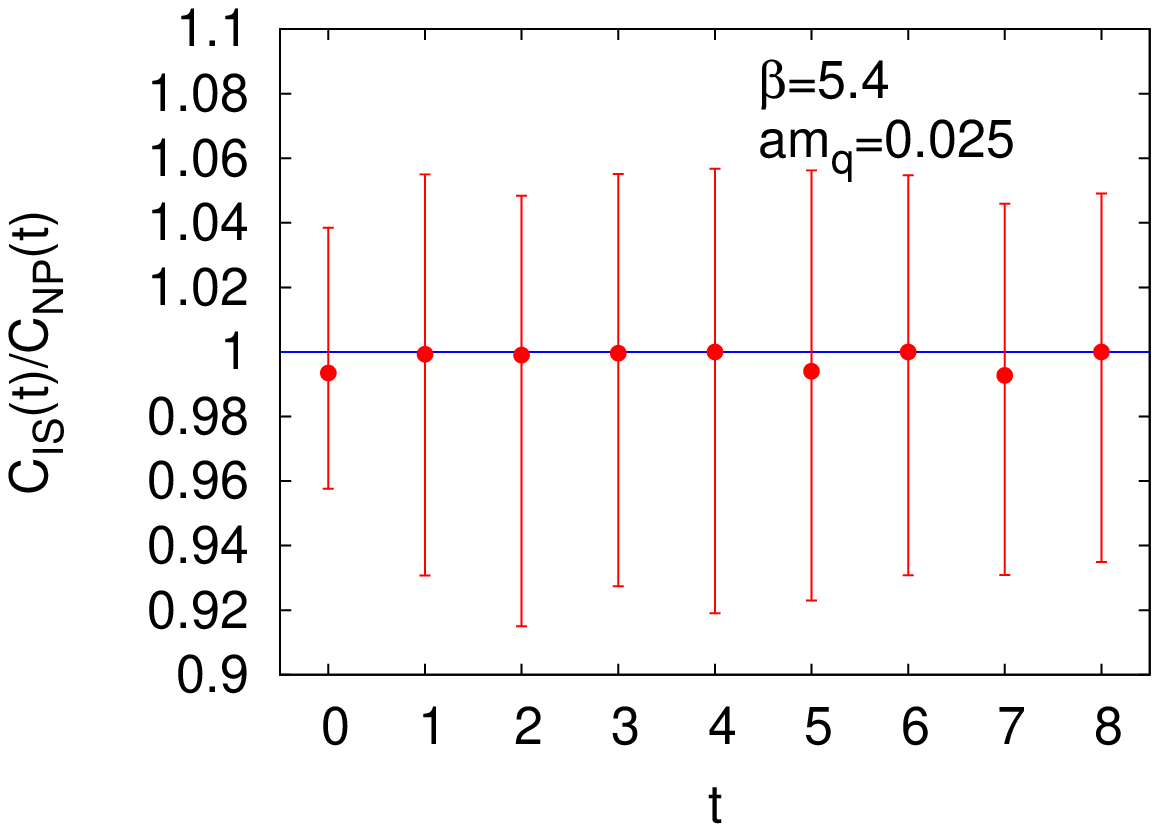}
\includegraphics[scale=0.7]{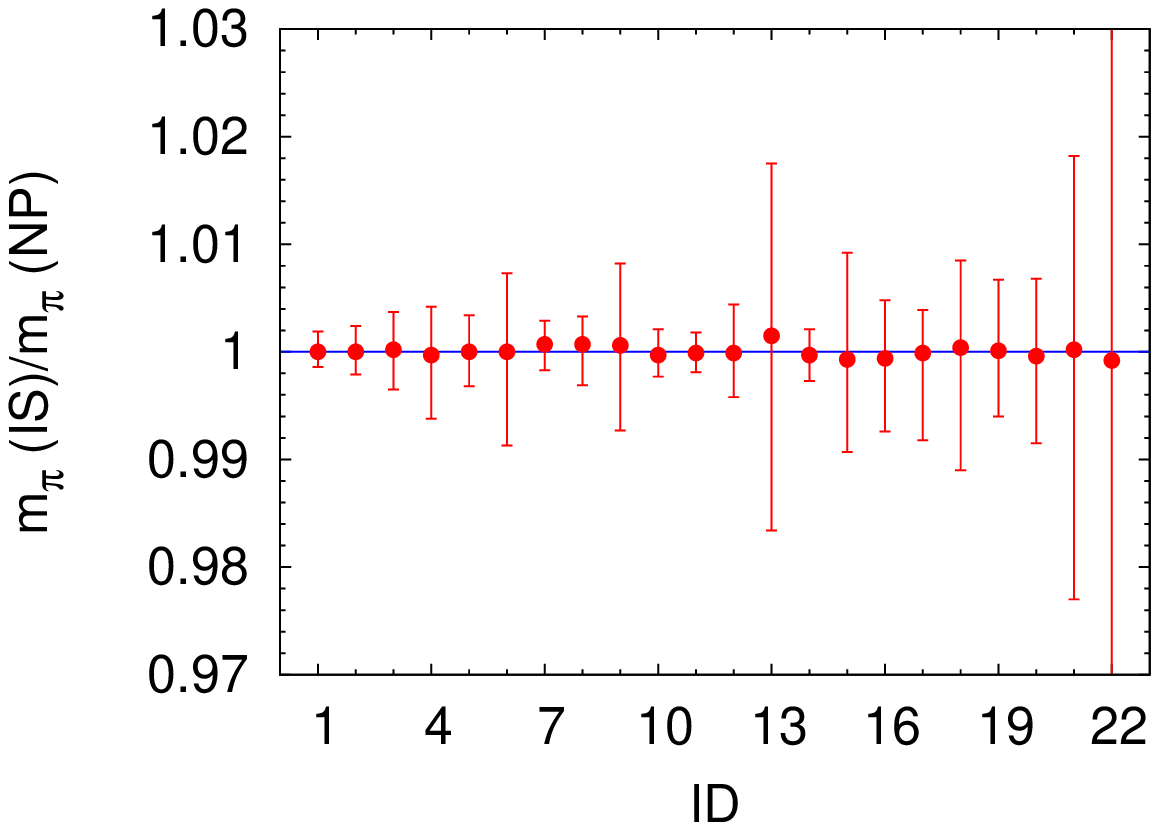}
\end{center}
\caption{The first panel shows the ratio of pion correlators extracted
 with independent bootstrap estimators of the correlator at each $t$
 (labelled IS) and using the same bootstrap configurations at all $t$
 (labelled NP, as before) at a representative run ID 6. The second panel
 shows the comparison of masses extracted from the two different estimators
 of the correlators.}
\eef{psisnp}

\bet[!htb]
\begin{center}
\begin{tabular}{c || c | c | c | c || c | c}
\hline
ID & $L/a$ & $\beta$ & $am_q$ & S & $am_\pi({\rm NP\/})$ & $am_\pi({\rm IS\/})$ \\
\hline
\hline
1 & 16 & 5.2875 & 0.1 & 50 & 0.7899 (27) & 0.7904 (23) \\
2 & 16 & 5.2875 & 0.05 & 50 & 0.5753 (23) & 0.5756 (27) \\
3 & 16 & 5.2875 & 0.025 & 70 & 0.4159 (26) & 0.4161 (27) \\
4 & 16 & 5.2875 & 0.015 & 50 & 0.3241 (24) & 0.3240 (34) \\
\hline
5 & 16 & 5.4 & 0.05 & 75 & 0.6033 (47) & 0.6030 (54) \\
6 & 16 & 5.4 & 0.025 & 51 & 0.4376 (61) & 0.4376 (71) \\
7 & 24 & 5.4 & 0.015 & 51 & 0.3500 (18) & 0.3504 (21) \\
8 & 32 & 5.4 & 0.01 & 40 & 0.2922 (17) & 0.2925 (12) \\
\hline
9 & 16 & 5.5 & 0.05 & 50 & 0.6184 (91) & 0.6177 (93) \\
10 & 24 & 5.5 & 0.025 & 101 & 0.4463 (22) & 0.4459 (22) \\
11 & 28 & 5.5 & 0.015 & 120 & 0.3542 (19) & 0.3541 (21) \\
12 & 32 & 5.5 & 0.01 & 40 & 0.2896 (21) & 0.2894 (30) \\
13 & 32 & 5.5 & 0.005 & 50 & 0.2129 (39) & 0.2124 (43) \\
\hline
14 & 24 & 5.6 & 0.05 & 55 & 0.5938 (29) & 0.5935 (27) \\
15 & 24 & 5.6 & 0.025 & 103 & 0.4255 (66) & 0.4246 (78) \\
16 & 28 & 5.6 & 0.015 & 120 & 0.3299 (49) & 0.3302 (56) \\
17 & 32 & 5.6 & 0.01 & 40 & 0.2682 (42) & 0.2685 (44) \\
18 & 32 & 5.6 & 0.005 & 50 & 0.1973 (45) & 0.1965 (30) \\
19 & 32 & 5.6 & 0.003 & 105 & 0.1506 (25) & 0.1513 (24) \\
\hline
20 & 24 & 5.7 & 0.025 & 59 & 0.3954 (73) & 0.3956 (59) \\
21 & 32 & 5.7 & 0.005 & 50 & 0.1751 (57) & 0.1738 (55) \\
22 & 32 & 5.7 & 0.003 & 50 & 0.134 (13) & 0.134 (19) \\
\hline
\end{tabular}
\end{center}
\caption{Comparison of pion masses extracted by the two methods labelled NP
 and IS, whose explanations are given in the text.}
\eet{psmass}

The molecular dynamics time separation between successive stored
configurations which we used are larger than those used before for naive
staggered quark simulations \cite{covmat, milc38, milc42, col67}, and
are typical of those used today. Additive increase of the MD time
separation decreases autocorrelations exponentially. So it is worthwhile
performing the analysis in which the correlation function at each distance
is re-sampled independently.

Such a comparison is shown in the first panel of \fgn{psisnp} for the
PS correlator in one of our simulations. The ratio of the correlation
function obtained through independent sampling (labelled IS) and that
obtained using our usual sampling (labelled NP as before) is completely
consistent with unity at all separations.  This conclusion is true
for all sets of simulations we have made.  A comparison of the masses
obtained by the two methods is shown in \fgn{psisnp} through the ratio
of the masses. As expected from the behaviour shown in the first panel,
the masses are not changed by the sampling of the correlator.  The pion
masses estimated using these two methods are collected in \tbn{psmass}.
From the Table one also sees that our exploration of statistical methods
covered a very wide range of pion masses in lattice units.

\section{The vector}\label{sec:rho}

\bef
\begin{center}
\includegraphics[scale=0.7]{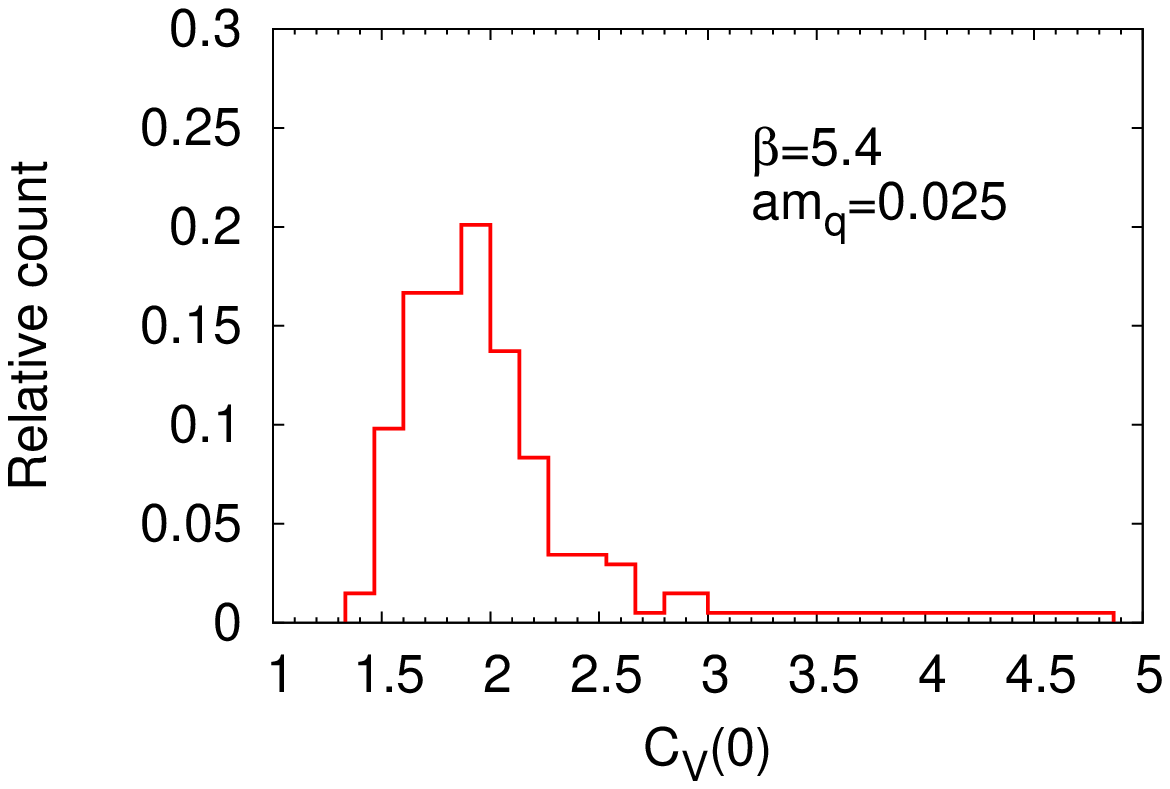}
\includegraphics[scale=0.7]{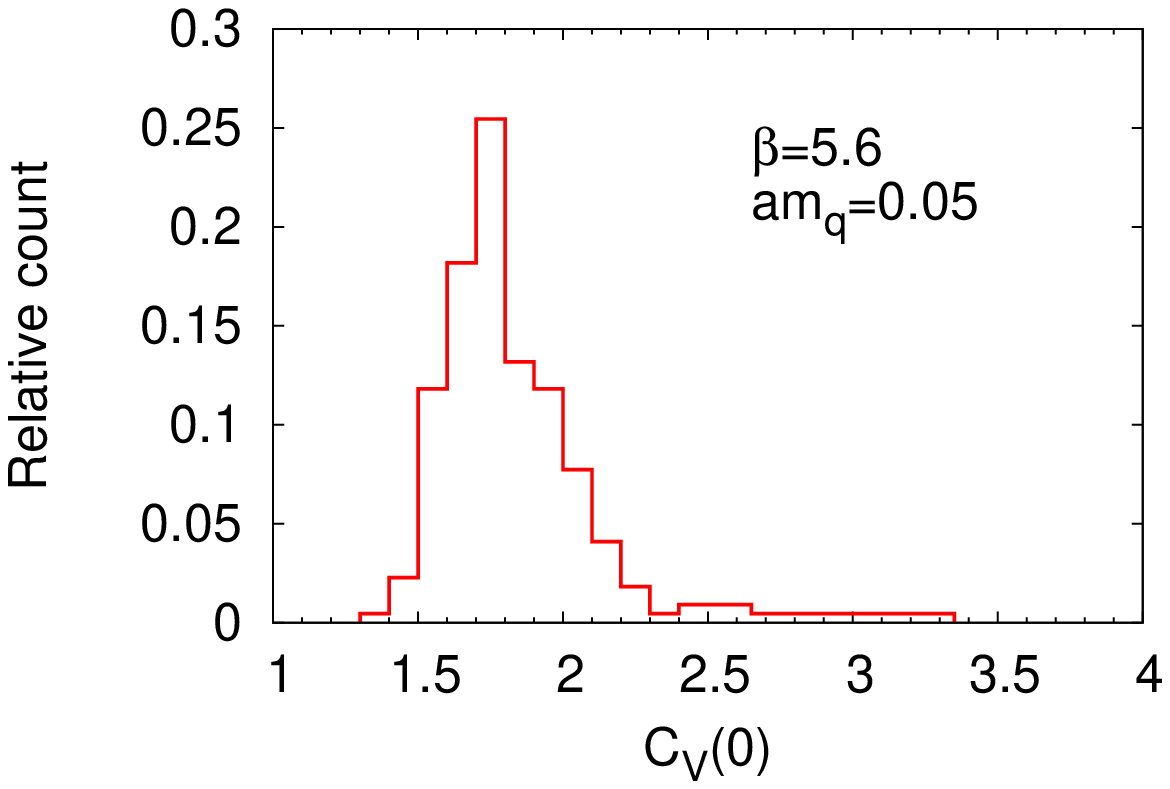}
\end{center}
\caption{The distribution of measurements of the local vector correlators
 at distance zero for the representative run IDs 6 and 14 seem to be
 non-Gaussian.  The configurations which give rise to the measurements
 at the tail of these distributions are generally those which populate
 the tail of the PS correlator.}
\eef{corrdist}

\bef
\begin{center}
\includegraphics[scale=0.7]{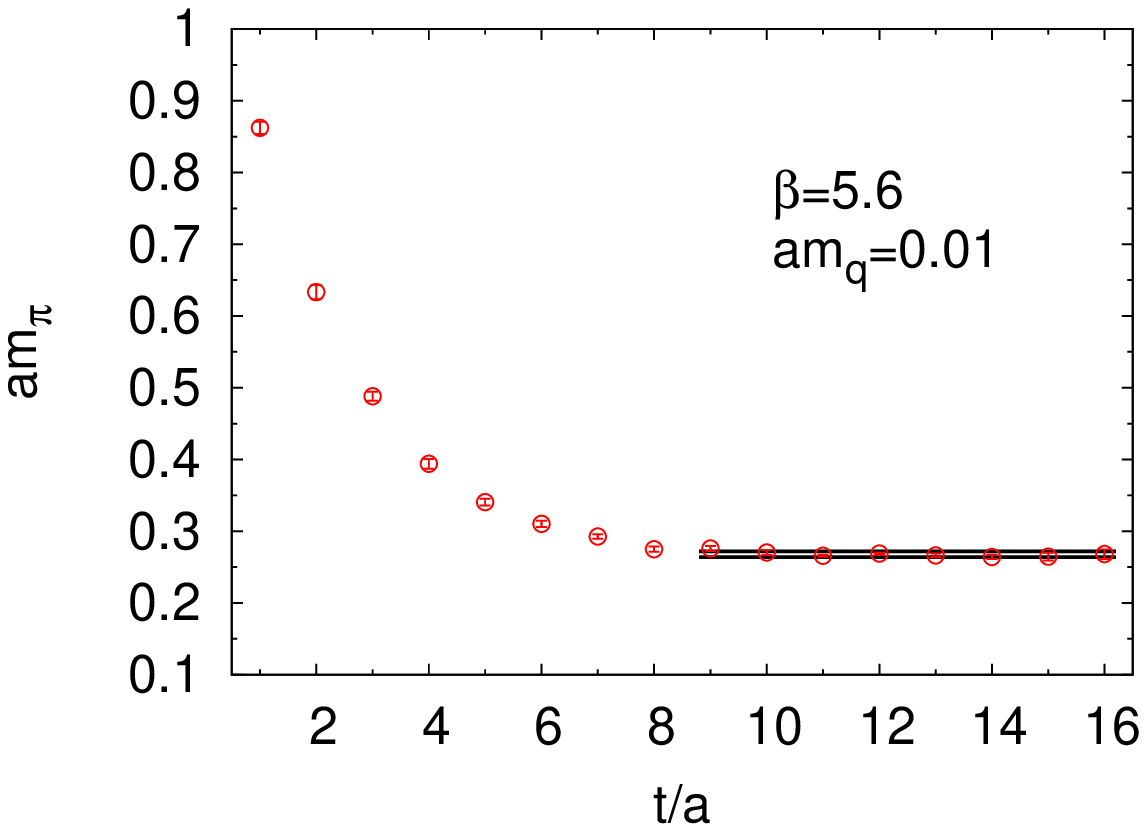}
\includegraphics[scale=0.7]{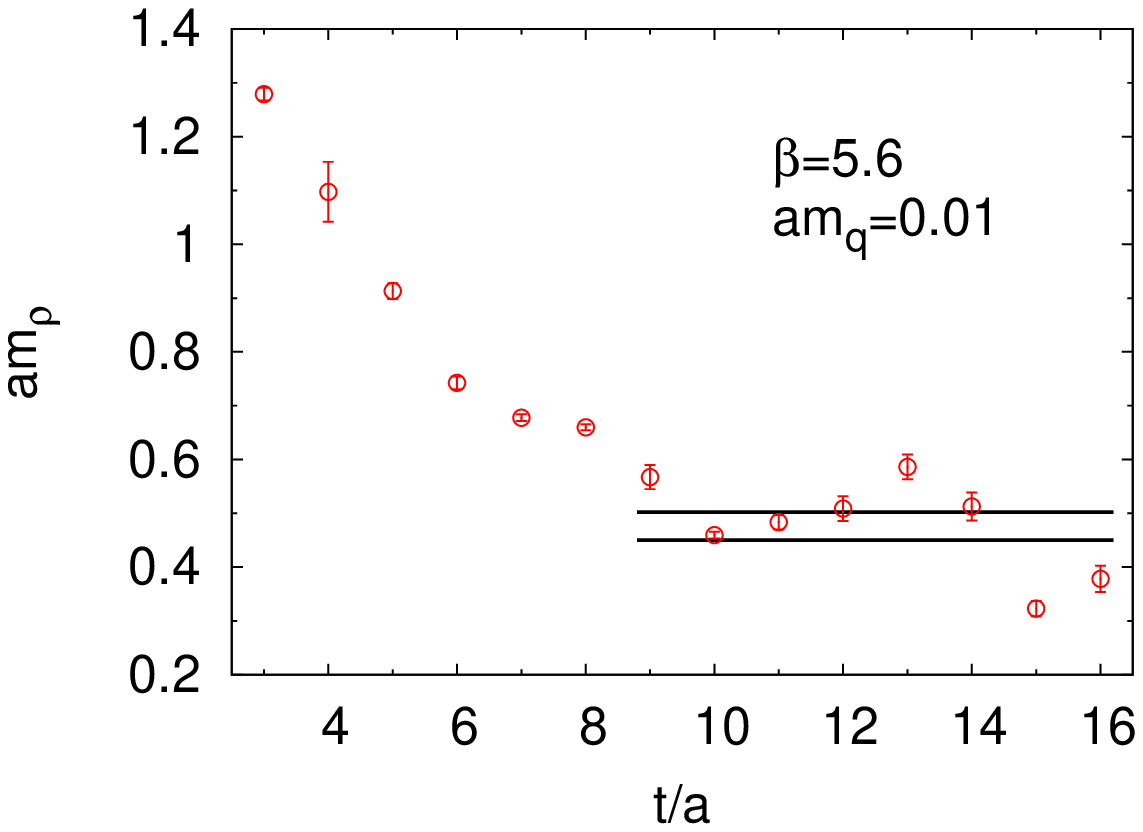}
\includegraphics[scale=0.7]{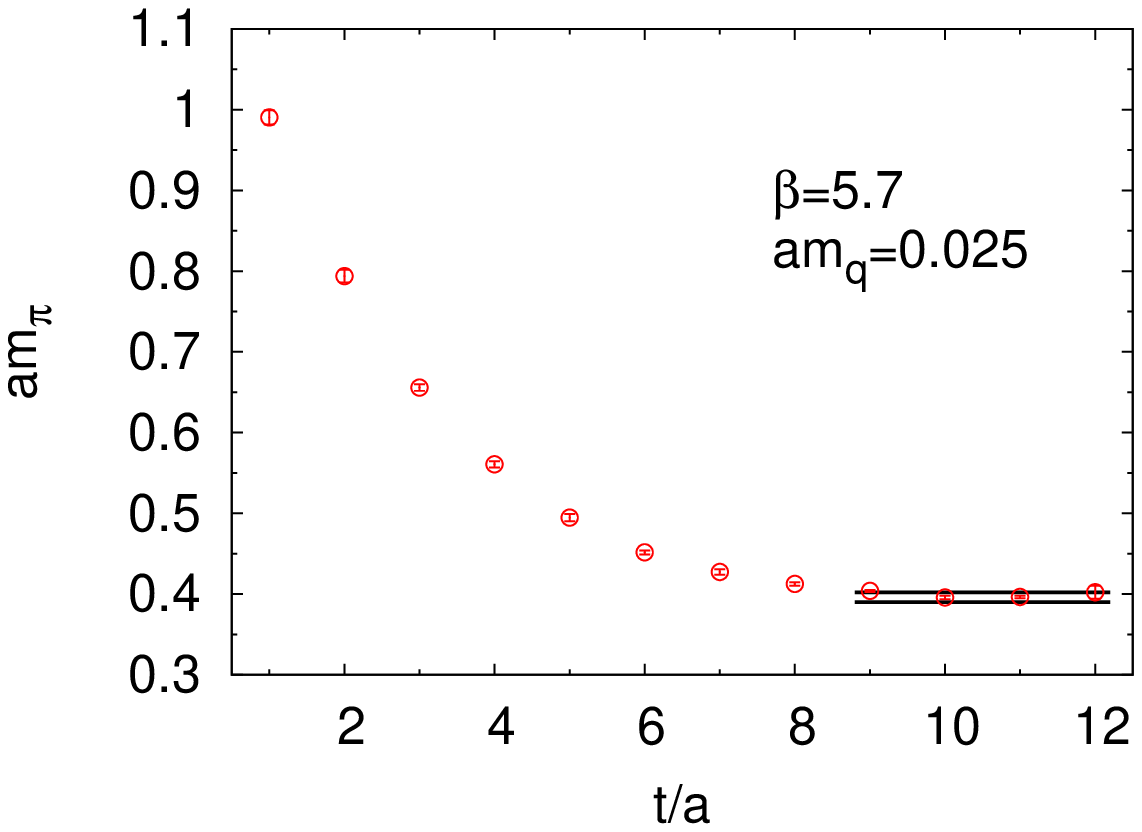}
\includegraphics[scale=0.7]{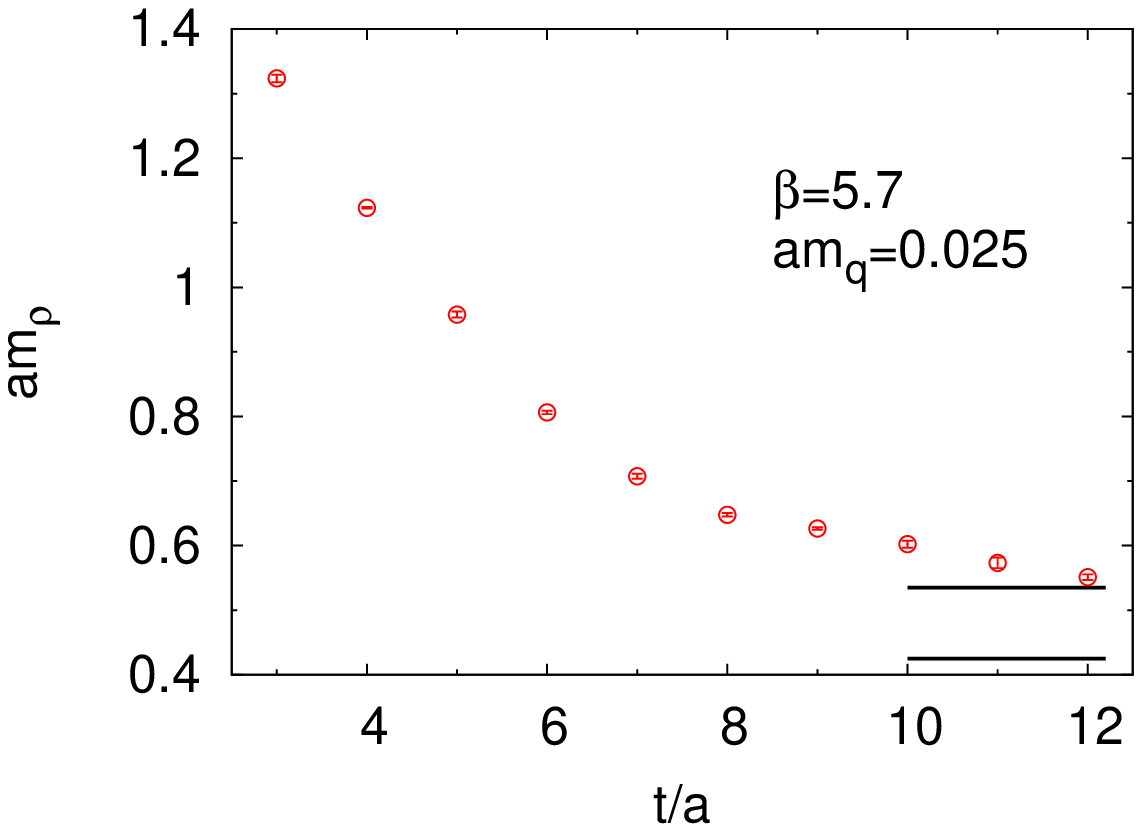}
\end{center}
\caption{Local masses for pions and vector mesons in some representative
 runs. The best fit masses are shown along with the local masses. For pions
 the band shows the 68\% statistical uncertainty band. For the vector meson the
 band is due to the systematic uncertainties, which happen to be larger than the
 statistical uncertainty. For $\beta=5.7$, $am_\rho$ is extracted after taking
 into account the excited state contribution.}
\eef{localm}

\bet[!htb]
\begin{center}
\begin{tabular}{c||c|c||c|c||c|c|}
\hline
ID&$\beta$&$am_q$&\multicolumn{2}{c||}{Single mass fit}&\multicolumn{2}{c|}{Two masses fit}\\
\hline
  & & & $am_\rho({\rm NP\/})$&$am_\rho({\rm IS\/})$&$am_\rho({\rm NP\/})$&$am_\rho({\rm IS\/})$\\
\hline
 1&5.2875&0.1&	1.464 (7)&	1.462 (8)& & \\
 2&5.2875&0.05&	1.336 (16)&	1.340 (12)& & \\
 3&5.2875&0.025& 1.289 (6)&	1.288 (6)& & \\
\hline
 5&5.4&0.05&	1.286 (4)&	1.286 (3)& & \\
 6&5.4&0.025&	1.177 (9)&	1.177 (12)& & \\
 7&5.4&0.015&	1.118 (6)&	1.117 (6)& & \\
\hline
 9&5.5&0.05&	1.046 (9)&	1.044 (9)& & \\
10&5.5&0.025&	0.904 (4)&	0.904 (4)& & \\
\hline
14&5.6&0.05&	0.844 (4)&	0.844 (5) &0.820 (19) (4) &0.819 (17) (6) \\
15&5.6&0.025&	0.638 (2)&	0.639 (4) &0.572 (16) (71)&0.565 (28) (79) \\
16&5.6&0.015&	0.595 (4)&	0.595 (4) &0.577 (15) (14)&0.571 (11) (26) \\
17&5.6&0.01&	0.475 (14)&	0.476 (13)& & \\
18&5.6&0.005&	0.420 (18)&	0.415 (18)&0.277 (39) (24)&0.291 (53) (24) \\
\hline
20&5.7&0.025&	0.568 (3)&	0.569 (5) &0.482 (28) (53)&0.480 (26) (55) \\
22&5.7&0.003&	0.418 (11)&	0.417 (8) &0.306 (36) (53)&0.304 (28) (72) \\
\hline
\end{tabular}
\end{center}
\caption{Comparison of vector masses extracted by the methods NP and IS,
 for various ID numbers. The IDs and the labels on masses have the same
 meanings as in \tbn{psmass}. In several cases the plateau in the local
 masses was too short to trust the single mass staggered fits, and the
 first excited state (and its staggered partner) was added to the fit.
 Data sets excluded from the table are too noisy for a stable fit.}
\eet{vmass}

\bef
\begin{center}
\includegraphics[scale=1.0]{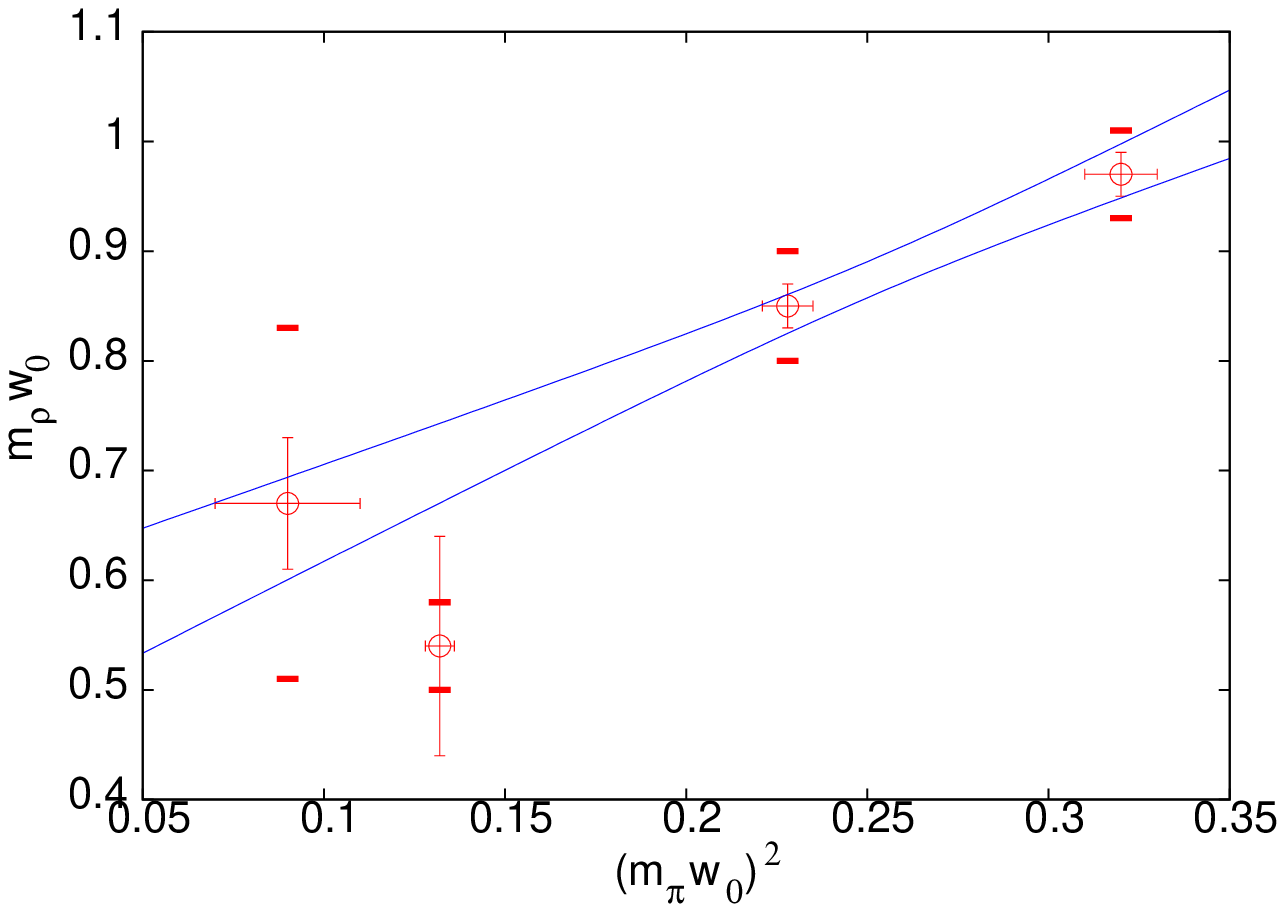}
\end{center}
\caption{$m_\rho w_0$ as a function of $(m_\pi w_0)^2$. The fat bars
 show systematic uncertainties; thin for statistical uncertainty. The
 blue lines show the 68\% uncertainty band of a linear extrapolation
 fitted to the four data points shown here.}
\eef{mpimrho}

On examining the distributions of other meson correlation functions, we
found that they are skewed in general. We show examples for the vector
correlator in \fgn{corrdist}. The tails of these distributions generally
come from the same configurations as for the pion. As before, we take
account of this skewness by using the methods, NP and IS, explored in
the previous section.

The analysis of these correlation functions is more complex than that of
the pseudo-Goldstone pion because the correlator contains an oscillating
staggered piece
\beq
   C_V(t) = A\cosh\left[m\left(\frac{aN_t}2-t\right)\right]
        + (-1)^{t/a} A'\cosh\left[m'\left(\frac{aN_t}2-t\right)\right],
\eeq{cstag}
where $a$ is the lattice spacing. We extract effective masses by using
four successive values of $t$ to extract the four parameters. We looked
for plateaus in such effective masses, and fitted the correlation function
above within the range of the plateau.  The statistical uncertainty in the
fitted mass, $ma$, was obtained using a pair of nested bootstraps. For
each set of re-samplings of the correlation function, the fit returned
a value of the parameters being fitted. A further bootstrap over this
process gave a distribution of the parameters. This was used to estimate
parameter uncertainty in the usual way. There are also systematic uncertainties
in the measurement.  The major uncertainty comes from having to choose the
range of the fit.  We quote a systematic uncertainty on the fitted parameters
as the maximum difference in the estimator of the mass when varying the
end points of fit range by one lattice unit.

As one can see from the local masses in \fgn{localm}, the region
over which a fit can be performed is significantly shorter for the
vector meson than for the pion.  The lack of perfect overlap with the
ground state limits the smallest values of $t$ which we can use in
the fit. As is visible in the figures, this is not necessarily a more
stringent cutoff for vectors than for pions. However, since the vector
mass is larger, the correlator falls faster, and the signal to noise
ratio deteriorates, limiting the largest $t$ which we can use. This 
large-$t$ problem can be beaten only with statistics, and therefore
represents a hard CPU limitation. Instead, as is common in the literature,
we try to use the small-$t$ information.  Since the correlator couples
to a tower of states, it gets a contribution in the form in \eqn{cstag}
from each state.  Techniques for extracting ground and excited states
simultaneously have been explored since long \cite{excited}. Notable
developments are the use of variational methods with many different
forms of the source \cite{variational} and Bayesian fitting \cite{bayes}.

Here we use the technically simpler alternative method which uses a fit
of two (four, with staggering) masses to the effective masses. Since
the number of fit parameters doubles, one has to take a sufficiently
large interval in $t$ to obtain a statistical test of the goodness of
fit. Clearly, there are two sources of systematic uncertainties: first
in determining the range over which the fit is performed, and second
in choosing the $t$ with which to associate the effective mass which
is being fitted. As before, we vary the fit interval by one unit at
each end point. We also let the effective mass be associated with every
separation which was used to determine it. The maximum variation in $ma$
obtained with these changes is quoted as a systematic uncertainty in this
method. 

Run IDs 3, 15 and 20 of \tbn{psmass} and \tbn{vmass} can be compared
with previous measurements of masses reported in the literature
\cite{milc38,milc42,col67}.  In all three cases we find good agreement
of previously quoted values of $am_\pi$ with the results we report
in \tbn{psmass}. We also find that the previously reported results on
$am_\rho$ are in reasonable agreement with the vector masses we extract
with single mass fits \tbn{vmass}.

Since we have previously determined the Wilson flow scale $w_0/a$, we
can examine our results for $am_\rho$ also in terms of $m_\rho w_0$.
In \fgn{mpimrho} we plot data on $m_\rho w_0$ as a function of $m_\pi
w_0$ for the four smallest pion masses we used at the finest lattices
possible. The coarsest lattice spacing among these corresponds
to $a<w_0/1.7$.  Since $m_\pi^2$ is proportional to a renormalized
quark mass, when this is small enough, $m_\rho$ should be linear in
this. Although the errors at the smallest quark masses are rather large,
the data seems to fall the range where the linear extrapolation seems
reasonable. In view of this, we fitted an extrapolation function for
$m_\rho w_0$ linear in terms of $(m_\pi w_0)^2$.  The 68\% uncertainty
band on this extrapolation are also shown in \fgn{mpimrho}.

The slope of the extrapolation may be
captured in the quantity
\beq
  S_\rho = 2\;\frac{m_\rho w_0(m_\pi w_0=0.4) - m_\rho w_0(m_\pi w_0=0.3)}
         {m_\rho w_0(m_\pi w_0=0.4) + m_\rho w_0(m_\pi w_0=0.3)}.
\eeq{sloperho}
The fit shown in \fgn{mpimrho} gives $S_\rho=0.14\pm0.08$. With our
set of simulations we are unable to remark on the possible lattice
spacing dependence of this slope. A previous estimate of the ratio
$w_0/\sqrt{t_0}$ with this set of simulations showed that a smooth limit
is reached at $a\simeq w_0/2$ \cite{prev}. While this could be taken as
an indication that the slope parameter we have determined is close to
its continuum value, it would be useful to check this in future.

With this fit we can extrapolate self-consistently to the physical
$m_\pi$, and there use the physical value of $m_\rho$ to extract $w_0$ in
physical units. We quote a statistical uncertainty in the extrapolation;
this arises from the statistical uncertainty in the mass measurements. We
also quote a systematic uncertainty which is the maximum difference
between this extrapolation and the two obtained by leaving out of the
fits either the measurement at the smallest or the largest $m_\pi$.
This gives $w_0=0.14\pm0.02\pm0.01$ fm, where the first uncertainty is
statistical and the second is systematic.  This extraction of the scale
$w_0$ is subject to the same caveats as the computations of the slope
parameter $S_\rho$.

\section{Conclusions}

\bet
\begin{tabular}{l|l|l||c|c||c||l|l|l|l|l}
\hline
$\beta$ & $ma$ & $N_s$ & Machine & Traj & Statistics 
 & $w_0/a$ & $am_\pi$ & $am_\rho$ & $m_\pi w_0$ & $m_\rho w_0$ \\
 & & $L/a$ & & (MD) & $T_0+T \times N$ & & & & & \\
\hline
5.2875	& 0.1	& 16 & V & 1 & $400+10\times50$ & 0.6112 (4) & 0.790 (2) & 1.462 (8) (7)  & 0.483 (1)  & 0.894 (5) (4) \\
	& 0.05	& 16 & V & 1 & $780+10\times50$ & 0.6354 (6) & 0.576 (3) & 1.340 (12) (7) & 0.366 (2)  & 0.851 (8) (4) \\
	& 0.025	& 16 & V & 1 & $200+15\times70$ & 0.6539 (1) & 0.416 (3) & 1.288 (6) (--) & 0.2714 (13) & 0.842 (4) (--) \\
	& 0.015	& 16 & V & 1 & $400+10\times50$ & 0.6608 (5) & 0.324 (3) &    ---         & 0.214 (2) & --- \\
\hline
5.4	& 0.05	& 16 & V & 2 & $200+20\times75$ & 0.8418 (14) & 0.603 (5) & 1.286 (3) (2)    & 0.508 (4) & 1.082 (3) (2) \\
	& 0.025 & 16 & V & 1 & $400+10\times51$ & 0.9264 (21) & 0.438 (7) & 1.177 (12) (23)  & 0.406 (6) & 1.09 (1) (2) \\
	& 0.015	& 24 & V & 2 & $400+10\times50$ & 0.9600 (9)  & 0.354 (2) & 1.117 (6) (10)   & 0.340 (2) & 1.072 (6) (10) \\
	& 0.01	& 32 & G & 2 & $200+20\times40$ & 0.9922 (7)  & 0.292 (1) &     ---          & 0.290 (1) & --- \\
\hline
5.5	& 0.05	& 16 & V & 1 & $200+20\times50$   & 1.1689 (40) & 0.618 (9) &     1.044 (9) (88) & 0.72 (1) & 1.22 (1) (10) \\
	& 0.025	& 24 & V & 1 & $1680+10\times101$ & 1.2651 (18) & 0.446 (2) &     0.904 (4) (11) & 0.564 (3) & 1.144 (5) (14) \\
	& 0.015	& 28 & G & 2 & $400+10\times120$  & 1.3302 (13) & 0.354 (2) &     ---   & 0.471 (3) & --- \\
	& 0.01	& 32 & G & 2 & $200+20\times40$   & 1.3771 (16) & 0.289 (3) &     ---   & 0.398 (4) & --- \\
	& 0.005	& 32 & BG & 1 & $250+10\times50$  & 1.4254 (37) & 0.212 (4) &     ---   & 0.302 (6) & --- \\
\hline
5.6	& 0.05	& 24 & V & 1 & $400+10\times55$  & 1.4850 (26) & 0.594 (3) & 0.819 (17) (6)   & 0.882 (5) & 1.22 (2) (1) \\
	& 0.025	& 24 & V & 1 & $1700+10\times48$ & 1.6007 (33) & 0.425 (8) & 0.565 (28) (79)  & 0.68 (1) & 0.90 (4) (13) \\
	& 0.015	& 28 & G & 2 & $400+10\times120$ & 1.7087 (25) & 0.330 (6) & 0.571 (11) (26)  & 0.56 (1) & 0.97 (2) (4) \\
	& 0.01	& 32 & G & 2 & $200+20\times40$  & 1.7814 (36) & 0.268 (4) & 0.476 (13) (26) & 0.477 (7) & 0.85 (2) (5) \\
	& 0.005	& 32 & BG & 1 & $300+10\times50$ & 1.8547 (71) & 0.196 (3) & 0.291 (53) (24) & 0.363 (6) & 0.54 (10) (4) \\
	& 0.003	& 32 & BG & 1 & $600+5\times105$ & 1.8824 (32) & 0.151 (2) &     ---    & 0.284 (4) & --- \\
\hline
5.7	& 0.025	& 24 & V & 1 & $530+10\times59$  & 1.9645 (48)  & 0.396 (5) & 0.480 (26) (55)  & 0.78 (1)  & 0.94 (5) (11) \\
	& 0.005	& 32 & BG & 1 & $370+10\times50$ & 2.1470 (73)  & 0.174 (5) &    ---          & 0.37 (1)  &   ---          \\
	& 0.003	& 32 & BG & 1 & $300+10\times50$ & 2.2103 (162) & 0.13 (2) & 0.304 (28) (72) & 0.30 (4) & 0.67 (6) (16) \\
\hline
\end{tabular}
\caption{A summary of our scale setting measurements. The scale setting by
 $w_0$ was reported in \cite{prev}. We have updated the pion masses presented
 in that study by reporting here the results of the analysis technique IS.
 For $m_\rho$ the statistical uncertainty are given before the systematic uncertainty.}
\eet{configs}

In \scn{pion} we have reported extensively on the statistical analysis
of masses. We observed that the distributions of correlation functions
are strongly skewed, and a Gaussian analysis of the sample can not
be justified.  We found that bootstrap estimates, which do not assume
any particular form of the distribution function, of the correlation
functions and their errors give sensible results.  We are not aware
of earlier systematic reports on the distribution of measurements
of correlation functions. We found that the masses obtained through
independent bootstrap sampling of the correlator at each $t$ (called
IS here) give results in complete agreement with sampling at all $t$
together (which we called NP). This is shown by the detailed compilation
of results in \tbn{psmass} and \tbn{vmass}.

The basic results we report in this paper are collected in \tbn{configs}.
The measurements of $w_0$ were reported in \cite{prev}; they are included
here for completeness. The pion and rho masses reported here are obtained
using the IS sampling technique. Where older results \cite{covmat,
milc38, milc42, col67} are available, they agree with ours within
statistical errors. Our results cover a wider range of lattice spacing and
renormalized quark mass than was available for $N_f=2$ naive staggered
quarks earlier. Our best estimate of the slope of the rho mass with the
pion mass is
\beq
   S_\rho = 0.14\pm0.08
\eeq{slopevalue}
where $S_\rho$ is defined in \eqn{sloperho}.

Since we have earlier determined the scale $w_0/a$ at these bare
parameters, we can now use these mass measurements to estimate the scale
$w_0$ from the vector meson mass.  Extrapolation to the physical pion mass
yields the value 
\beq
   w_0=0.14\pm0.02\pm0.01 {\rm\ fm}, 
\eeq{w0rho}
where the first uncertainty is due to statistical uncertainties in
the extraction of masses, and the second from an estimate of the
uncertainty in the extrapolation to physical pion mass.  This value
for $N_f=2$ naive staggered quarks should be compared to the value
$w_0=0.13^{+0.01}_{-0.02}$ fm, obtained earlier by a comparison with
scale setting using the Lepage-Mackenzie method for the extraction of
$\Lambda_\MSbar$ \cite{prev}. Since the earlier extraction could require
better control of UV artifacts than available at present, the current
extraction, using a long-distance measurement, is an useful alternative
method. It would be interesting to verify this estimate using other
scales in future.

We also point out that an extraction of $w_0$ for $N_f=2$ clover quarks
using $f_K$ to set the scale yields a larger value, namely $w_0=0.1757\pm
0.0013$ fm, where the uncertainty is statistical \cite{bruno}. While
the difference is not large or statistically very significant, it is
intriguing. Tracing the source of this difference will certainly allow us
to understand the continuum limits of various fermion measurements better.

\appendix
\section{Distribution of sample medians}\label{sec:apone}

Suppose $r$ is a real random variate with distribution $f(r)$, and let $\mu$
be the population median. The cumulative distribution of $r$ is
\beq
  F(r) = \int_{-\infty}^r f(t) dt,
\eeq{cum}
where $F(\infty)=1$, $F(-\infty)=0$, and $F(\mu)=1/2$. If we draw $2n+1$
samples from the population and find
that the median of the sample is $z$, then the probability of this being
so is
\beq
  g(z) = {{2n+1}\choose n} \left\{F(z)[1-F(z)]\right\}^n f(z).
\eeq{sample}
When
$n$ is large enough, we expect $z$ to be close enough to $\mu$ so that
the Taylor expansion
\beq
  F(z) = \frac12 + f(\mu) (z-\mu) + \frac12 f'(\mu)(z-\mu)^2 + \cdots
\eeq{taylor}
converges sufficiently quickly. Then, using Stirling's approximation in
the binomial coefficient, we find that
\beq
  \log g(z) \simeq -4n[f(\mu)]^2(z-\mu)^2 + \cdots
\eeq{gauss}
This proves that $g(z)$ is Gaussian with mean $\mu$ and variance of
$1/(8n[f(\mu)]^2)$. The error in the estimate of $\mu$ therefore
decreases as $1/\sqrt n$. This result is attributed to Laplace
\cite{stigler}. The proof given here is an adaptation of one from
\cite{millers}.

This construction fails when $f(\mu)=0$. Then, retaining the next term
in the expansion, one can prove that $g(z)$ is even narrower. Such
constructions fail completely when $f(r)$ vanishes identically in a
region around $r=\mu$, so that a Taylor expansion of \eqn{taylor} is
impossible. However, this class of probability densities is
different than that for which the central limit theorem fails. An instructive
example with such a pathology is
\beq
   f(r) = \frac12\left[\delta(r)+\delta(r-1)\right].
\eeq{bimodal}

\end{document}